\documentclass[aps,prb,twocolumn,showpacs]{revtex4}
\usepackage{graphicx}

\begin{document}

\title{Rectangular quantum dots in high magnetic fields}

\author{E.~R\"as\"anen}
\email[Electronic address: ]{ehr@fyslab.hut.fi}
\author{A.~Harju}
\author{M.~J.~Puska}
\author{R.~M.~Nieminen}
\address{Laboratory of Physics, Helsinki University of Technology,
P.O. Box 1100, FIN-02015 HUT, FINLAND}

\date{\today}

\begin{abstract}
We use density-functional methods to study the effects of an 
external magnetic field on two-dimensional quantum dots with a 
rectangular hard-wall confining potential.
The increasing magnetic field
leads to spin polarization and formation of a highly inhomogeneous 
maximum-density droplet at the predicted magnetic field strength.
At higher fields, we find an oscillating behavior in the electron
density and in the magnetization of the dot. 
We identify a rich variety of phenomena behind the
periodicity and analyze the complicated many-electron
dynamics, which is shown to be highly dependent on the shape of
the quantum dot. 
\end{abstract}


\maketitle

\section{Introduction}

The problem of many interacting electrons in a non-circular 
quantum well under the influence of an external magnetic field 
is extremely challenging for the computational tools available. 
For the time being, theoretical studies 
of such systems have dealt either with very small
electron numbers, or with the single-electron properties 
in non-integrable quantum billiards.~\cite{stockmann} 
On the other hand, the rapid technical development in the 
realization of different quantum-dot systems naturally 
motivates theoretical modeling of complicated 
many-electron structures.~\cite{revmod}

The magnetic-field dependence
of the energy spectrum in a square two-electron quantum dot has 
been analyzed in detail by Creffield {\em et al.}~\cite{creffield2}
They found Aharonov-Bohm-type oscillations in the lowest levels,
indicating periodic singlet-triplet changes in the ground state.
Ugajin~\cite{ugajin} found that these state transitions lead 
to strong effects in 
the optical excitation spectra. Within a similar square-dot 
system, he studied the effects of the Coulomb interaction on the 
far-infrared-absorption (FIR) spectra.~\cite{ugajin2}
Recent density-functional FIR calculations for soft-wall triangular
and square dots have been done by Val\'{\i}n-Rodr\'{\i}quez 
{\em et al.},~\cite{rodri} who identified corner and side modes
in the system. The interactions and the quantum-dot geometry
affect also strongly the magnetization, at least for small electron 
numbers.~\cite{magnusdottir,sheng}

In a circular geometry, a maximum-density droplet (MDD) 
caused by the magnetic field corresponds 
to a polarized state with electrons occupying successive angular 
momentum states, i.e., $l=0,-1,...,-N+1$, giving
$L_{\rm MDD}=-\frac{1}{2}N(N-1)$ for the total angular momentum of 
$N$ electrons. 
A stability region for the MDD has been identified 
in electron transport experiments through vertical
quantum dots by Oosterkamp {\em et al.}~\cite{oosterkamp}
This state, as well as the post-MDD regime in higher magnetic 
fields, has been considered theoretically in circular
parabolically-confined quantum dots by several authors.~\cite{ferconi,mddreimann,ari} 
In the case of a non-circular symmetry, however, the high-magnetic-field
limit has not been previously analyzed to the best of our knowledge.

In this work, we examine the magnetic-field effects in rectangular
quantum dots by using the spin-density-functional theory (SDFT).
Our aim is to define the 
MDD state in a rectangular geometry and extend our previous analysis 
of the MDD formation in hard-wall quantum dots.~\cite{physica}
The onset of the MDD can be predicted
from the number of flux quanta which only depends on the area of
the dot and is thus geometry-independent. We study also
the beyond-MDD regime that can be characterized by
periodic state oscillations. The periodicity can be observed in
the magnetization which we compare with the corresponding results for
noninteracting electron systems. The origin of the periodicity is 
generally the
competition between magnetic confinement and Coulomb repulsion,
but the behavior of the single-electron states is very 
sensitive to the dot geometry.

This paper is organized as follows.
In Sec.~\ref{sec2} we present briefly the model Hamiltonian and 
the computational methods, based on a real-space spin-density-functional 
approach. In Sec.~\ref{sec4} we consider the formation and the structure
of the MDD in rectangular geometries.
Section~\ref{sec5} presents the state oscillations above the MDD regime, 
first from the point of view of the total magnetization, and finally 
at the level of the effective single-electron states. 
The paper is summarized in Sec.~\ref{sec6}.

\section{The model and the methods} \label{sec2}

We restrict the dot to the {\em xy} plane and
use the effective mass approximation with the material parameters for
GaAs, i.e., the effective mass $m^*$=0.067 $m_e$ and the dielectric constant
$\epsilon=12.7$. The many-body Hamiltonian reads
\begin{eqnarray}
H & = & \frac{1}{2m^*}\sum^N_{i=1}\left[-i\hbar\nabla_i+\frac{e}{c}\mathbf{A}
({\mathbf r}_i)\right]^2
+\sum^N_{i<j}\frac{e^2}{\epsilon|{\mathbf r}_i-{\mathbf r}_j|} \nonumber \\
& + & \sum^N_{i=1}\left[V_{\rm ext}({\mathbf r}_i)+g^*\mu_BBs_{z,i}\right],
\label{hami}
\end{eqnarray}
where the shape of the dot is defined by a rectangular hard-wall
confining potential,
\begin{equation}
V_{\rm ext}(x,y)=\left\{ \begin{array}{ll}
0, & 0\leq{x}\leq\beta{L},\,0\leq{y}\leq{L}\\
\infty, & \text{ otherwise}.
\end{array} \right.
\end{equation}
Here $\beta$ defines the ratio of the side legths of the rectangle, and
$L$ is scaled such that the area of the dot remains constant, i.e.,
$\beta{L^2}=4\pi^2\,{a^*_B}^2$, where $1\,a^*_B\approx{10.03}\,{\rm nm}$.
We apply an external magnetic field ${\mathbf B}=B\hat{z}$ 
perpendicular on the {\em xy} plane, and use the symmetric gauge,
\begin{equation}
\mathbf{A}=\frac{B}{2}(-y,x,0),
\end{equation}
for the vector potential. The last term in Eq.~(\ref{hami}) is
the Zeeman energy (with $g^*=-0.44$ for GaAs)
which has an insignificant effect on the results presented and is thus
neglected. In the calculations we employ the SDFT
in the self-consistent Kohn-Sham (KS) formulation. In the systems 
considered in this study,
the SDFT gives reasonably accurate results compared to the 
computationally more demanding current-spin-density-functional 
approach. In the local spin-density approximation we use
the exchange-correlation energy by Attaccalite 
{\em et al.} \cite{attaccalite} It is based on the
diffusion Monte Carlo simulations over the whole range 
of spin polarization, which is a major improvement
comparing to the previous parametrizations. This leads to more 
accurate results in both zero and non-zero magnetic fields
as shown in Ref.~\onlinecite{lsda}.

The calculations are perfomed in real space using
finite differences for the derivative operations on 
two-dimensional point grids.
Since there are no implicit restrictions for the symmetry,
the external potential can be shaped arbitrarily in
the computing region. The number of grid points is
128 $\times$ 128, giving a sufficient accuracy in the 
total energy. To make the convergence effective,
we employ the Rayleigh quotient multigrid
method \cite{mika} for the discretized single-electron 
Schr\"odinger equations. 

\section{MDD formation} \label{sec4}

In the zero-field solution for a rectangular quantum dot
with an even number of electrons,
the total spin is either zero or
one, depending on the degeneracy on the highest level.~\cite{recta}
Increasing $B$ brings the higher spin states closer to
the $S=0$ and $S=1$ states in energy, and finally the full spin 
polarization is found. The value of $B$ for the complete spin
polarization is rather independent of the deformation
parameter.~\cite{physica} In the eight-electron dot, for example,
the ground state becomes fully spin-polarized at $B\sim 10$ T
as $\beta=1$ and at $11$ T as $\beta=2$. 

\begin{figure}
\includegraphics[width=8.5cm]{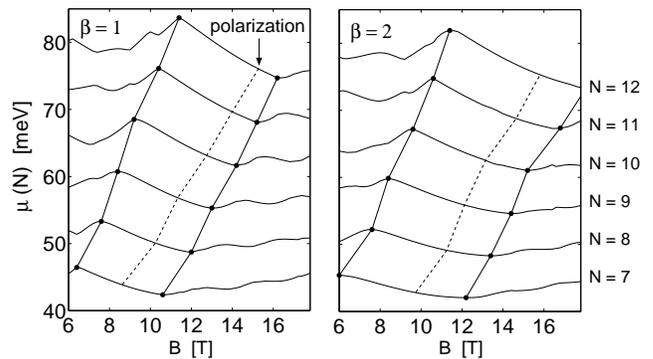}
\caption{Chemical potentials of the fully polarized states for
$N$-electron rectangular quantum dots as a function of the magnetic 
field $B$.}
\label{chems}
\end{figure}

Figure~\ref{chems} shows the chemical potentials,
$\mu(N)=E(N)-E(N-1)$, of the \emph{polarized} states as a
function of $B$ for $N=7-12$. 
The point when the totally polarized state becomes the 
ground state is marked with a dotted line in the figure. 
We consider only the polarized state in order to clarify
the MDD window, which corresponds to the descending regime
in the chemical potential. Similar behavior in $\mu$ was
observed in the experiments for vertical quantum dots by 
by Oosterkamp {\em et al.}~\cite{oosterkamp}. 
They measured
the evolution of the Coulomb blockade peaks as a function of $B$
for $N=0-40$ and found clearly identifiable phases for the
filling factor, of which $\nu=1$ corresponds to a maximum-density
droplet. According to our definition, the true MDD
exists on the rightmost ascending stripe in Fig.~\ref{chems}. 
As $\beta$ increases, the MDD
windows become larger and flatter due to pronounced
localization in the {\em x} direction.
Thus, the square case shows 
the most drastic behavior because of the highest consistency
of the dot symmetry with the magnetic confinement.

\begin{figure}
\includegraphics[width=7.5cm]{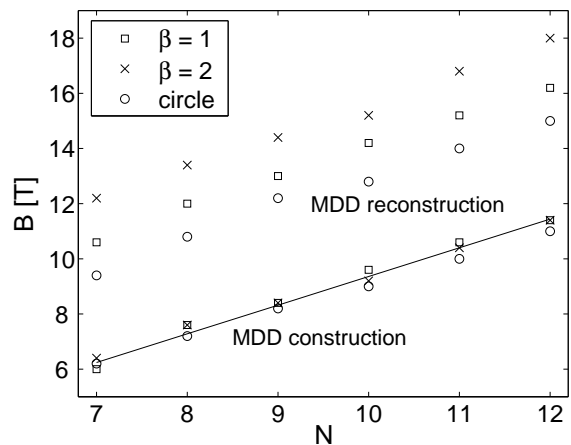}
\caption{MDD-window limits as a function of the number of electrons
in different quantum-dot geometries. The line
for the predicted MDD formation is also shown.}
\label{borders}
\end{figure}

Figure~\ref{borders} gives the limits for the MDD window
as a function of $N$ in different quantum-dot geometries.
We see that the increase in the width of the MDD window
is due to a shift in the reconstruction point, whereas the
onset of the window is nearly independent of the dot geometry.
The reason is the following. The formation of the MDD requires
an equal number of flux quanta, $N_\Phi=\Phi/\Phi_0$, to the $N-1$ 
vortices 'seen' by an electron as the other electrons are fixed.
A vortex is defined as a point where the wave function
is zero and its phase changes by $-2\pi$ for a counterclockwise
rotation.~\cite{pekka}
Here $\Phi_0=h/e$ is the single flux quantum, 
and since the magnetic flux is directly proportional to the dot area,
i.e., $\Phi=BA$, we thus get $B=h(N-1)/eA\approx 1.04(N-1)$ T for 
the onset of the MDD in all geometries. We show this prediction in 
Fig.~\ref{borders}
and it agrees remarkably well with the transition points obtained from
the chemical potentials. The reconstruction limit is, however, much more
difficult to estimate and it depends strongly on the geometry as
explained above. 

\begin{figure}
\includegraphics[width=8.5cm]{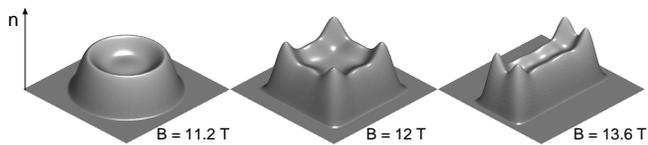}
\caption{Electron densities for the MDD states
of eight-electron quantum dots with different geometries.}
\label{mdds}
\end{figure}

At the end of the MDD window the droplet achieves the
smoothest electron density.
In Fig.~\ref{mdds} we show the maximum-density
distributions for $\beta=1$ and $\beta=2$ quantum dots, and also
for a circular hard-wall dot with $N=8$. 
They have a highly inhomogeneous structure
instead of a flat distribution found in parabolically confined 
MDD's.~\cite{mddreimann} This is a direct consequence of
the hard-wall confinement where the Coulomb interaction favors
strong localization in the corners and on the edges
over the whole range of $B$.
In the circular hard-wall MDD, the occupation of the 
angular momentum orbitals follows the
original definition of a maximum-density droplet for a 
parabolic dot, i.e., $l_z=0,-1,...,-7$. The number of 
vortices in a particular KS state thus equals $|l_z|$. 
For $N$ electrons at the positions $z_i=x_i+i y_i$,
the many-body wave function for a parabolic
MDD can be written in Laughlin's~\cite{laughlin} form as
\begin{equation}
\Psi_{\rm MDD}=\exp\left(-\frac{\pi}{2}{B}\Phi_0\sum_k|z_k|^2\right)\prod_{i<j}^N(z_i-z_j),
\end{equation}
which is similar to the wave function for
a state with the filling factor $\nu=1$ in the thermodynamic limit.
It is easy to see that by following the above definition for a vortex
we obtain the correct total number of $N-1$ vortices in the system. 

The rectangular dot shows a more complicated
structure: the formation of the MDD corresponds to the
appearance of the correct maximum number of $N-1$ vortices in the 
KS state highest in energy, but in the other states there 
are typically 'extra' vortices, leading to more
than $N(N-1)/2$ of them in the KS states altogether. 
The reason is the complex
structure of the indefinite angular momentum states that prevents
the description of the non-circular MDD with a Laughlin-type 
many-body wave function, even if the $B$-$N$ dependence for 
the MDD formation can be correctly estimated.
Thus, we resort to considering the evolution of the 
expectation values of the angular momentum operator,
$\hat{l}_z=-i\hbar(x\frac{\partial}{\partial{y}}-y\frac{\partial}
{\partial{x}})$, for different KS states in the high-$B$ regime.
Below we designate this expectation value for non-circular dots
as the effective $l_z$.

\section{State oscillations} \label{sec5}

The rapidly growing Coulomb interaction leads to
the reconstruction of the MDD into a lower-density droplet 
with a higher angular momentum as $B$ increases.
According to our calculations, the collapse of the circular 
hard-wall MDD leads into a ringlike structure with 
$l_z=-1,-2,...,-N$, contrary to the mean-field results 
for parabolic dots, where the Chamon-Wen edge states
and charge-density waves have been found.~\cite{mddreimann}
Due to the hard-wall confinement in our case,
there is a remarkably higher density on the edge already 
in the MDD state, and the exchange-correlation energy is thus relatively 
less important than in a parabolic dot. However, exact 
diagonalization and quantum Monte Carlo results predict 
circularly symmetric reconstruction also in the parabolic 
case.~\cite{ari} In a rectangular geometry, the result is a 
strongly localized edge state that resembles a Wigner molecule
in the sense that there are pronounced peaks in the electron
density to minimize the Coulomb interaction.~\cite{poly}

As $B$ is increased further, the electronic structure begins to oscillate 
between well-localized and more diffuse states. This is a result from 
the interplay between the magnetic confinement and the Coulomb interaction. 
In this connection, the polarized ground state changes periodically as the 
interactions entangle the single-electron states on the lowest Landau level. 
This is similar to singlet-triplet oscillations found in interacting 
two-electron quantum dots of both circular~\cite{wagner} and 
square~\cite{creffield2} shapes, as well as in a double-dot 
system.~\cite{ari2}

\subsection{Magnetization} \label{sec5a}

\begin{figure}
\includegraphics[width=7.5cm]{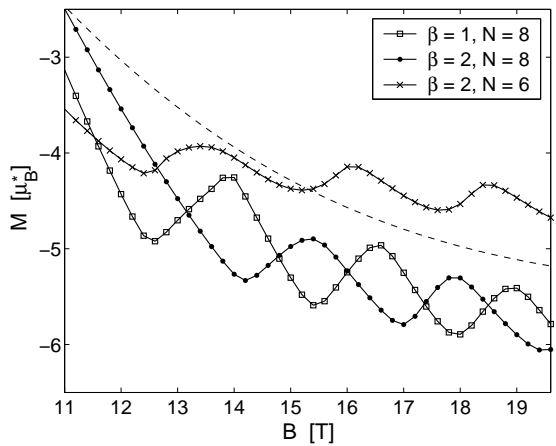}
\caption{Magnetization in three different rectangular quantum-dot
systems beyond the MDD regime. The dashed line indicates the noninteracting
case of ($\beta,N$) = (2,6). The effective Bohr magneton 
$\mu_B^*=e\hbar/2m^*$.}
\label{magnet}
\end{figure}

The structural oscillations beyond the MDD state can be observed in 
the magnetization
of the dot, since the growth in the total angular momentum 
leads to the decrease in the current.
The total magnetization for a finite system is defined at 
zero temperature as
\begin{equation}
M=-\frac{\partial E_{\rm tot}}{\partial B}.
\end{equation}
In Fig.~\ref{magnet} we show the magnetization in three rectangular quantum
dots beyond the MDD regime. The local minima correspond to the
points where the localization on the edge is maximized, and
at local maxima there are states with a density contribution
at the center. It can be clearly seen in Fig.~\ref{magnet} that
$M$ decreases with increasing electron number, and the onset of
the oscillation moves to higher $B$ as the deformation $\beta$ is
increased, which is consistent with the upper limit of the MDD 
window shown in Fig.~\ref{borders}.
The period in magnetic field strength is nearly a constant, $\sim 2.5$ T
in all the systems presented in Fig.~\ref{magnet}. We emphasize
that the majority of ($\beta,N$) combinations lead to more complicated
and even irregular magnetization. This will be discussed 
in more detail in Sec.~\ref{sec5b}. Figure~\ref{magnet} also
shows the noninteracting magnetization for ($\beta,N$) = (2,6), calculated
from the six lowest eigenenergies on the first Landau level.
In the MDD regime, it is $\sim 1\,\mu_B^*$ above the interacting 
curve with the same slope. At $B \sim 15$ T, however, there is a 
clear crossing which indicates that the interactions 
retard the overall increase in $|M|$ as a function on $B$ in 
the oscillatory regime.

Magnetization in square-shaped quantum dots up to $N=4$ has been 
studied by Magn\'usd\'ottir and 
Gudmundsson.~\cite{magnusdottir} They found remarkably higher 
magnetization in the noninteracting than in the interacting 
picture, but did not consider the high-$B$ limit, where we see 
the crossing between those two cases. In a two-electron square dot
studied by Sheng and Xu,~\cite{sheng}
the magnetization looks similar to our case, but the oscillation
represents the trivial singlet-triplet exchange.~\cite{sheng}

\subsection{Development of the angular momenta} \label{sec5b}

\begin{figure}
\includegraphics[width=5cm]{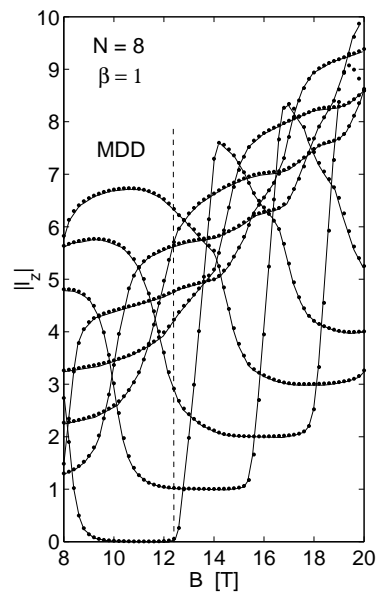}
\caption{Evolution of the effective $l_z$ values
as a function of the magnetic field $B$ in a square 
eight-electron quantum dot.}
\label{lznel}
\end{figure}

As a mean-field theory, the density-functional approach only
approximates the many-body energy states.
However, we will next consider the development of the 
effective angular momentum of different KS orbitals in order to obtain
a deeper insight on the state oscillations on the single-electron 
level. In Fig.~\ref{lznel}
we present the evolution of the effective $l_z$ values of
the occupied KS levels in the polarized state of a square 
eight-electron quantum dot as a function of $B$. 
The structure is intriguing and shows clearly two different
modes in the system. First, there are four closely entangled states,
representing the relatively stable corner modes.
They are reminiscent of the Aharonov-Bohm-like energy-level 
structure in an interacting two-electron dot.~\cite{creffield2} 
Second, there are states that with increasing $B$
condense to nearly circular orbits on the first 
Landau level before the Coulomb interaction expands them to high 
effective $l_z$ values. In this context, the state development is similar 
to the circular case, where the occupation shifts from the 
lowest $l_z$ state to the highest level as the ground state changes. 
The periodic spread of the Landau states is clearly the origin 
of the magnetization oscillations discussed above.
The upper limit of the MDD window, marked in Fig.~\ref{lznel}, 
can be interpreted as the most evenly extended effective $l_z$ 
distribution similar to the corresponding circularly symmetric system.

\begin{figure*}
\includegraphics[width=12cm]{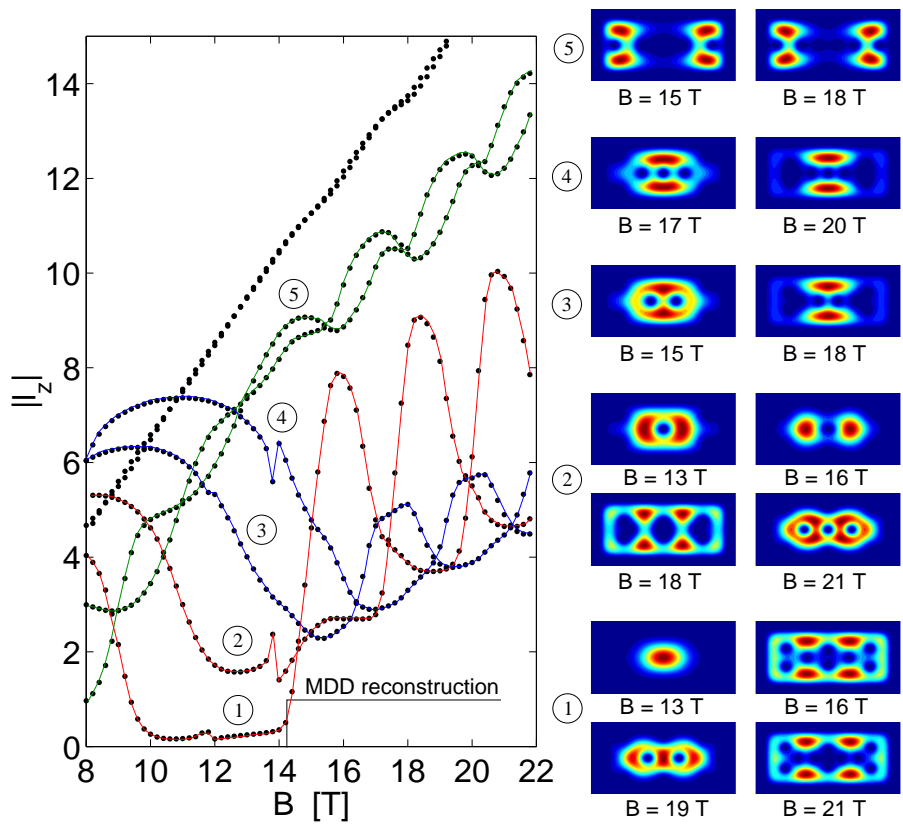}
\caption{Evolution of the effective $l_z$ values
as a function of the magnetic field $B$ in a $\beta=2$ 
eight-electron quantum dot. 
KS wave functions, corresponding to different modes, are 
also shown.}
\label{lzkai}
\end{figure*}

Figure~\ref{lzkai} 
shows the development of the effective $l_z$ values in the case of 
$\beta=2$. Due to the reduction of the symmetry, the evolution is 
more complicated than in the square case, and we can find four 
different modes. First, there are two corner modes in the system. 
Of them, the highest effective $l_z$ state shows steady localization 
at the both ends of the rectangle, and the second one (5) oscillates 
weakly. The latter is visualized in the right panel
of Fig.~\ref{lzkai} with the KS wave functions, 
$|\psi^i_{\rm KS}|^2$ (see also Ref.~\onlinecite{animation}).
In addition, there are two different types of highly oscillatory 
states. The first pair (3,4) has a remarkably lower
amplitude in the oscillation regime ($B\geq{14}$ T). As can be seen in 
the KS wave functions, they correspond to localization in the {\em y}
direction. Interlocked to these states, there are strongly oscillative
states (1,2) that, according to the KS wave functions, seem to form 
a periodic structure in the rectangle,
reminiscent of a 'lozenge' orbit in the correponding
classical bouncing map, studied in the square case by Aguiar.~\cite{aguiar}
It is noticeable that these states affect strongly on the corner modes
by compressing them slightly to lower $l_z$.
The cusps at $B=11.8$ T and $13.8$ T are numerical noise, corresponding 
to degeneracies of the KS orbitals 1 and 3 or 2 and 4, respectively.

It is clear from Fig.~\ref{lzkai} that the formation of bulk Landau 
states disappears as the
dot is squeezed toward a rectangular shape. 
Thus, the electron-electron interactions make the
quantum dot sensitive to the deformation parameter. On the other hand,
the overall behavior beyond the MDD is qualitatively similar 
in all geometries, even if the evolution of the effective $l_z$ values gets
very complicated. This occurs e.g. in the case of $\beta=1.5$, where
the different modes discussed above are hardly recognizable.
In this sense, an increase in the symmetry separates the different
states and makes the corresponding classically stable orbits visible.

We underline that within the density-functional theory,
one always has to regard carefully the KS states
and the corresponding wave functions shown in Figs.~\ref{lznel}
and \ref{lzkai}. In this particular system, however, we find 
that the procedure gives important information on the many-particle 
dynamics in an interacting non-integrable quantum dot, not accessible
with more 'exact' computational methods.

\section{Summary} \label{sec6}

To summarize, we have studied the electronic structure of rectangular
quantum dots in a magnetic field. We have identified the formation
of a maximum-density droplet at the predicted $B$, even if 
the electron density is strongly localized due to the hard-wall 
confinement. Beyond the MDD, we have found oscillating magnetization 
that reflects the periodic changes in the polarized ground state.
In the oscillating regime, the magnetization shifts to lower values 
than in the corresponding noninteracting case. The origin of the periodic 
behavior has been analyzed by following the evolution of the effective 
angular momentum quantum numbers. 
There are clear corner modes in all rectangular systems, 
but the development of the center orbits is found to be very sensitive 
to the geometry of the dot. In a square, the central KS orbitals are nearly
symmetric, whereas a rectangle with $\beta=2$ shows two different
oscillative modes. When the symmetry is reduced further, the center
orbitals become unindentifiable. In spite of the limitations in the 
mean-field approach, we have managed to bring out the richness of 
phenomena exhibited by the electronic structure in the high-$B$ limit.
We hope that this would inspire new experimental work on
this topic.

\begin{acknowledgments}
This research has been supported by the Academy of Finland through 
its Centers of Excellence Program (2000-2005). E. R. is also grateful
to Magnus Ehrnrooth foundation for financial support.
 
\end{acknowledgments}

\end{document}